\documentclass[twocolumn,english,aps,pre,showpacs,nofootinbib]{revtex4}
\usepackage[T1]{fontenc}
\usepackage[latin9]{inputenc}
\usepackage{amsmath}
\usepackage{graphicx}
\usepackage{amssymb}
\usepackage{esint}
\usepackage{nicefrac}
\usepackage{xfrac}
\usepackage[normalem]{ulem}
\usepackage[usenames]{color}

\begin{document}

\title{How the DNA sequence affects the Hill curve of transcriptional response}
\author{M. Sheinman}
\affiliation{Department of Physics and Astronomy, Vrije
Universiteit, Amsterdam, The Netherlands}
\author{Y. Kafri}
\affiliation{Department of Physics, Technion, Haifa 32000, Israel}

\date{\today}

\begin{abstract}
The Hill coefficient is often used as a direct measure of the cooperativity of  binding processes. It is an essential tool for probing properties of reactions in many biochemical systems. Here we analyze existing experimental data and demonstrate that the Hill coefficient characterizing the binding of transcription factors to their cognate sites can in fact be larger than one -- the standard indication of cooperativity -- even in the absence of any standard cooperative binding mechanism. By studying the problem analytically, we demonstrate that this effect occurs due to the disordered binding energy of the transcription factor to the DNA molecule and the steric interactions between the different copies of the transcription factor. We show that the enhanced Hill coefficient implies a significant reduction in the number of copies of the transcription factors which is needed to occupy a cognate site and, in many cases, can explain existing estimates for numbers of the transcription factors in cells. The mechanism is general and should be applicable to other biological recognition processes.
\end{abstract}
\maketitle

Molecular recognition plays an important role in biological systems ranging from antigen-antibody identification to protein-protein binding \cite{ABLRRW94}. In many cases the recognition process is driven by free-energy differences between a desired reaction and many competing undesired reactions \cite{Perelson97,Zhang2008Constraints,GMH2002,SM2004}.  One example, of particular importance, is that of protein-DNA interactions. Its role in understanding regulation in cells has led to large experimental effort to which aims at mapping the binding energy between transcription factors (TFs) in their specific state and different subsequences on the DNA \cite{Wang2011InVitro}. It is known that to a good approximation the energy can be written as a sum of energies representing the binding energy of a nucleotide on the DNA to the region on the protein with which it is aligned \cite{BH87,SF98}. Specifically, the binding energy of a nucleotide $s=A,C,G,T$ to position $j=1,2,...,L$ (where $L$ is the length of the protein's DNA binding domain in units of basepairs) is usually described by a $4 \times L$ position weight  matrix (PWM), $\epsilon_{s,j}$. By now, the PWM is known for many proteins and, together with a knowledge of the genomic sequence, it specifies the binding energy landscape of TFs to the DNA. 

Irrespective of the energy landscape properties the activation of a cognate site---a specific location on the DNA---by a TF is usually described by a Hill curve \cite{PKT2009}. Namely, if we consider a DNA molecular inside a container representing, say, a prokaryotic cell the activation probability of an operator by a TF is given by,
\begin{equation}
P_T=\frac{1}{1+\left( \frac{m_{1/2}}{m}\right)^n }.
\label{Hill}
\end{equation}
Here $m$ is the number of TFs in the cell and at $m=m_{1/2}$ the occupation probability is one half (the conversion to concentrations is trivial). The Hill coefficient (HC), $n$, governs the steepness of the curve and is widely used to extract qualitative information about the regulation of genes from experimental data \cite{Buchler2005Nonlinear,Kuhlman2007Combinatorial,Kim2009Transcriptional,Garcia2010Transcription}.  In the simplest cases, when there is no cooperative binding involved one expects $n=1$. In the presence of cooperative interactions $n$ is different than one. For example, in the case of activation by dimers one expects $n=2$ if $m$ is the number of monomers.

In this article we demonstrate that this simple intuitive picture for the Hill curve can fail. This is a direct consequence of a non-trivial combination of variations in the binding energy of TFs to different sites along the DNA and the steric repulsion between them. This leads to (a) a {\it disorder enhanced Hill coefficient} which is larger than one even in the absence of any cooperative binding to the operator, and (b) a dramatic increase in the occupation probability of the cognate site as compared to a system with no steric interactions between the TFs or a constant non-cognate binding energy. Importantly, we show that the results are essential for explaining the number of TFs found in cells.

\begin{figure}
\includegraphics[width=\columnwidth,trim=0 0 0 0,clip=true]{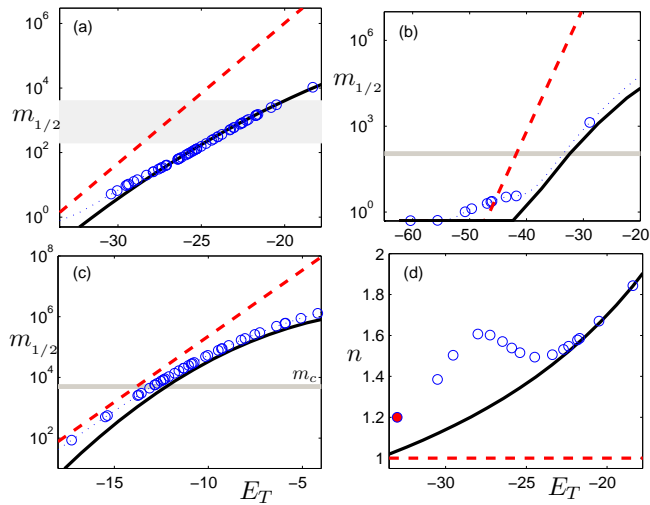}
\caption{The value of $m_{1/2}$ is presented as a function of the energy of different cognate sequences, $E_T$ for (a) LexA, (b) RpoN and (c) Lrp. The dotted lines are based on a numerical calculation using the \textit{E.coli} genome. The circles represent energies calculated from all known cognate sites of the TF. The black solid lines are based on the freezing regime approximation, Eq. \eqref{FreezeSat1}, while the red dashed lines are based on the non-steric approximation, Eq. \eqref{SelfAveSat1}.
Filled, grey horizontal areas show the typical range of the TF's copy number in \textit{E.coli}. The symbol $m_c$ in (c) marks the crossover value of the Lrp TF between the uncrowded and crowded regimes, predicted by Eqs. \eqref{nc1} and \eqref{nc2}. (d) The HC, obtained by a fit of the numerical data to Eq. \ref{Hill}, is shown as a function of the cognate site energy for the LexA TF. The solid line is the analytical prediction, Eq. \eqref{FreezeSat2}, while the circles represent numerical data based on real DNA and cognate sites sequence. The filled circle represents the HC of a hypothetical cognate site with a perfect consensus sequence. The dashed line represents the result of the non-steric approximation, Eq. \eqref{Annealed2}, which gives $n=1$.
\label{HalfOccNum}}
\end{figure}
\begin{figure*}
\includegraphics[width=\textwidth,trim=0 3 0 2,clip=true]{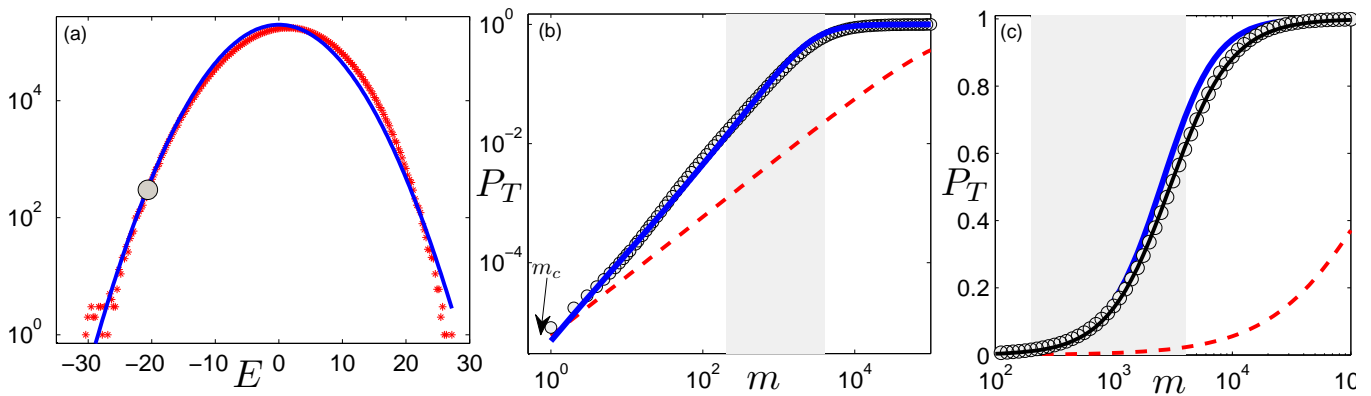}
\caption{(a) A histogram of the binding energies of the LexA TF to the \textit{E. coli} genome. The red points represent numerical data while the line is based on a Gaussian fit with $\sigma=5.76$. The big filled circle marks the binding energy to the recQ operon sequence with $E_T=-20.6$. (b) The occupation probability of the LexA protein to the recQ operon as a function of $m$. The circles represent numerical data based on the PMW of the protein and the \textit{E. coli} DNA sequence. The dashed line is based on the non-steric approximation, Eq. \eqref{Annealed}, while the solid line on Eq. \eqref{Freeze}. The solid arrow shows the crossover value (nonphysical in this case) of the protein copy number, $m_c=0.65$, predicted by Eqs. \eqref{nc1}, and \eqref{nc2}. (c) The same data as shown in panel (b) with a Hill function, Eq. \eqref{Hill}, fit (thin line) to the numerical data (circles) with $n=1.7$ and $m_{1/2}=3000$. The gray areas in (b) and (c) show the typical range of the LexA copy number in \textit{E.coli} ($200-4000$) \cite{Dri1994Control}.
\label{LexA}}
\end{figure*}
\begin{figure*}
\includegraphics[width=\textwidth,trim=0 10 0 5,clip=true]{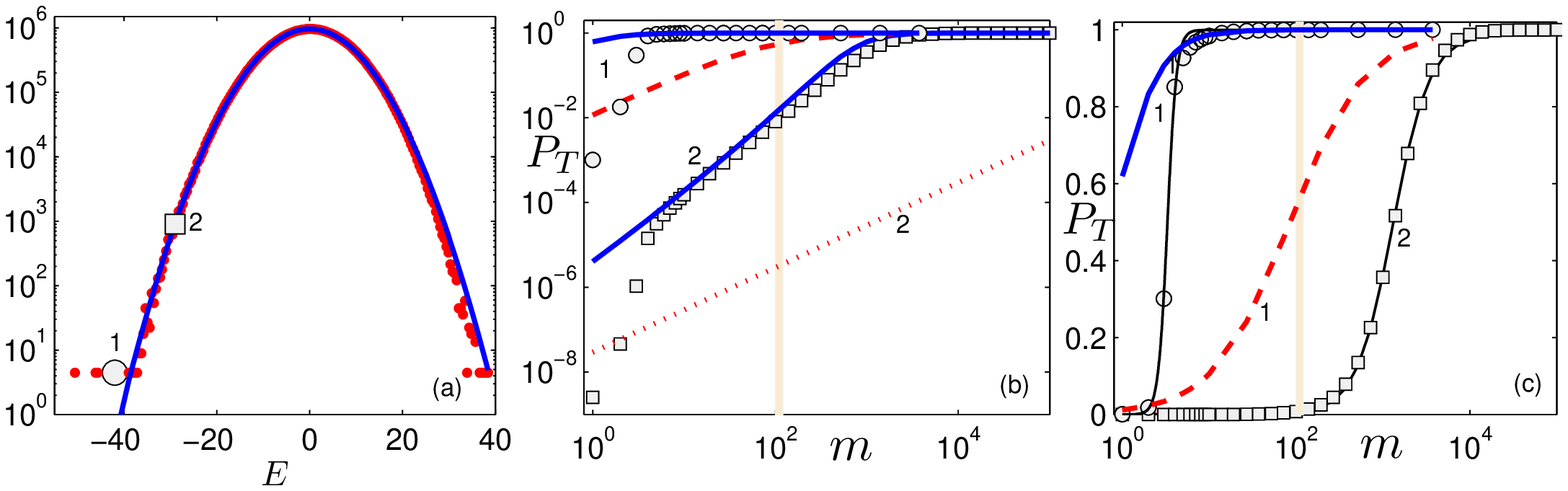}
\caption{(a) A histogram of the binding energy of the RpoN TF to the \textit{E. coli} genome. The red points represent the numerical data while the line is based on a Gaussian fit with $\sigma=7.8$. The circle shows the binding energy to the argT operon with $E_T=-41.8$. The square  shows the binding energy to the fixABCX operon with $E_T=-28.9$. (b) The occupation probability of the RpoN protein for the operons argT (circles and $1$ symbol) and fixABCX (squares and $ 2 $ symbol) as a function of $m$. The circles and squares represent numerical data based on the PWMs of the TFs and the \textit{E. coli} DNA sequence. The dashed (dotted) line is based on the non-steric approximation, Eq. \eqref{Annealed}, for the argT (fixABCX) operon, while the solid lines are based on Eq. \eqref{Freeze}. (c) The same numerical data as in panel (b) with a fit of a Hill function, Eq. \eqref{Hill}, (thin, solid, black lines) to the numerical data with $n=8.3$ and $m_{1/2}=3.3$ for the argT operon and with $n=2.07$ and $m_{1/2}=1300$ for the fixABCX operon. Vertical areas in panels (b) and (c) show the typical number of the RpoN proteins in \textit{E.coli} ($\sim 110$) \cite{Jishage1996Regulation}. 
\label{RpoN}}
\end{figure*}

\section{Results}
The Hill curve, Eq. \eqref{Hill}, is directly related to a formulation of the problem using statistical mechanics and the knowledge of the experimentally measured binding energy landscape of the TF to the DNA. To illustrate this we first focus  on a simple case where: \textit{(i)} There is no cooperativity associated with the structure of the TF or its binding properties, such that one would naively expects $n=1$. \textit{(ii)} The probability of the TF to be off the DNA  or in a non-specific conformation on the DNA is negligible. Note that by a non-specific conformation we mean one where the TF is on the DNA but does not interact with the bases. This conformation, which typically occurs due to electrostatic interactions, exists on any location along the DNA, including the cognate site. The effects of both simplifications are discussed in the Supplementary Information (SI). Using standard statistical mechanics it is straightforward to calculate $m_{1/2}$ and $P_{T}(m)$ numerically (see Methods). We use the PRODORIC database for PWMs of \textit{E.coli} TFs \cite{Munch2003PRODORIC}, their cognate sequences and the genomic sequence of \textit{E.coli} of length $N=2 \times 4686077$ \cite{EcoliGenome}. In the next section we discuss the results of this approach and show that it can lead to rather counterintuitive Hill curves. We then give a simple theory which accounts for the results analytically and gives precise conditions for these effects to be important.

\subsection{\textit{E.coli} TFs Hill curves}
First, we evaluate numerically $m_{1/2}$ for all known cognate sites of three TFs. The results are shown in Fig. \ref{HalfOccNum} (for the moment focus on the data shown as circles in panels $a, b$ and $c$). As can be seen (and intuitively clear), the weaker the binding energy of the cognate site (larger $E_T$) the larger $m_{1/2}$. The values of $m_{1/2}$ span a large range for different TFs and for different cognate sites of the same TF.
Next, in Figs. \ref{LexA} and \ref{RpoN} ($b$ and $ c $ panels) we present the occupation probability of typical cognate sites for two representative TFs, one cognate site of LexA and two of RpoN.  In addition we fit the results to a Hill curve. With the value of $m_{1/2}$ given the only fit parameter is the HC, $n$. Surprisingly, for the three cases (and others not shown) we obtain $n>1$ (in Figs. \ref{LexA} and \ref{RpoN} we obtain $n=1.7,\ 8.3,\ 2.03$), despite the absence of cooperativity in the model between the TF copies apart from steric interactions. As shown below this occurs for many TFs.

Remarkably, performing the same procedure but ignoring the steric repulsion of the TFs everywhere apart from the cognate site one obtains qualitatively and quantitatively distinct results (see Figs. \ref{HalfOccNum}, \ref{LexA} and \ref{RpoN}). This is despite the fact that in most cases the ratio of the number of TFs in the cell to the length of the DNA is very small. In fact, ignoring steric repulsion leads to a significant increasing in the value of $ m_{1/2} $ (in some cases much above the measured estimate of number of TFs in the cell). Moreover, this simplification {\it always} yields a HC of $n=1$.  Similarly, it is clear that without disorder in the binding energy of the TF to different sites along the DNA one would obtain $n=1$.

In sum, to account properly for the Hill curve of TFs binding to DNA one must (a) account for the disordered binding energy of the TF to different sites along thee DNA and (b) account for steric interactions between different copies of the TFs. Without these the HC would be lower than the actual one (later we show that this is true even for cases where naively one expects a HC which is larger than one) and $m_{1/2}$ would be much larger than the real one. Those are the main results of this paper. In what follows we explain the obtained results using an analytical approach. We show that the effect is generic and illustrate its importance for many TFs.

\subsection{Analytical solution}
As stated above, the binding energy to any sequence on the DNA is a sum of $L=10-30$ independent variables. Therefore, assuming a pseudorandom sequence on the non-cognate DNA, the probability distribution of the binding energy, $\Pr\left( E_i\right)$, is close to normal (see \cite{GMH2002,SM2004} and Figs. \ref{LexA}(a) and \ref{RpoN}(a)):
\begin{equation}
\Pr\left( E_i\right) \propto \exp\left( -\frac{E^2_i}{2\sigma^2}\right) \;.
\label{Pr}
\end{equation}
Here we define the centre of the normal distribution as zero energy.
The value of $ \sigma $ characterizes the width of the disordered binding energy profile along the DNA and is usually in the range $2-8$, where throughout the paper we measure energy in units of $k_B T$ with $k_B$ the Boltzmann constant and $T$ the temperature. The validity of this approximation is illustration in Figs. \ref{LexA}(a) and \ref{RpoN}(a) where we plot the histograms of the binding energies of two typical TFs on the \textit{E.coli} DNA.

\begin{figure}
\includegraphics[width=\columnwidth]{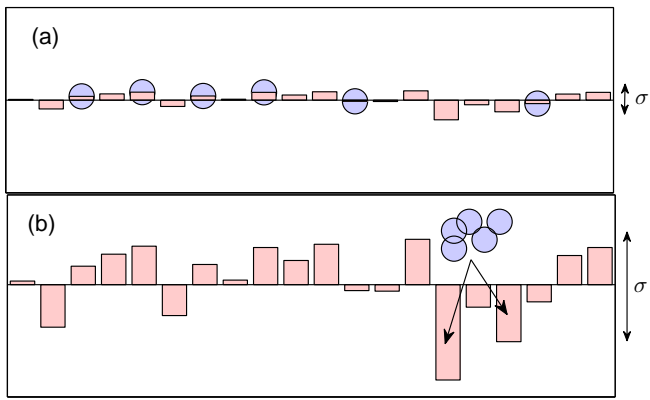}
\caption{An illustration of the crowding effect in  an energy landscapes with high disorder. Typical disordered energy profiles, with small and high disorder strength, are plotted in the (a) and (b) panels respectively. The circles represent TFs who's number is much smaller than the possible binding sites on the  DNA. Therefore, in the small disorder case, (a), they spread evenly across the DNA. When the disorder strength is high (panel (b)) the TFs compete on a limited number of traps. Then steric interaction between those competing on these small number of traps become important. As the traps fill with TFs the occupation probability of the cognate site increases dramatically. This leads to a reduction in $m_{1/2}$ and an enhanced HC.} 
\label{Scheme}
\end{figure}
\begin{figure*}
\includegraphics[width=\textwidth]{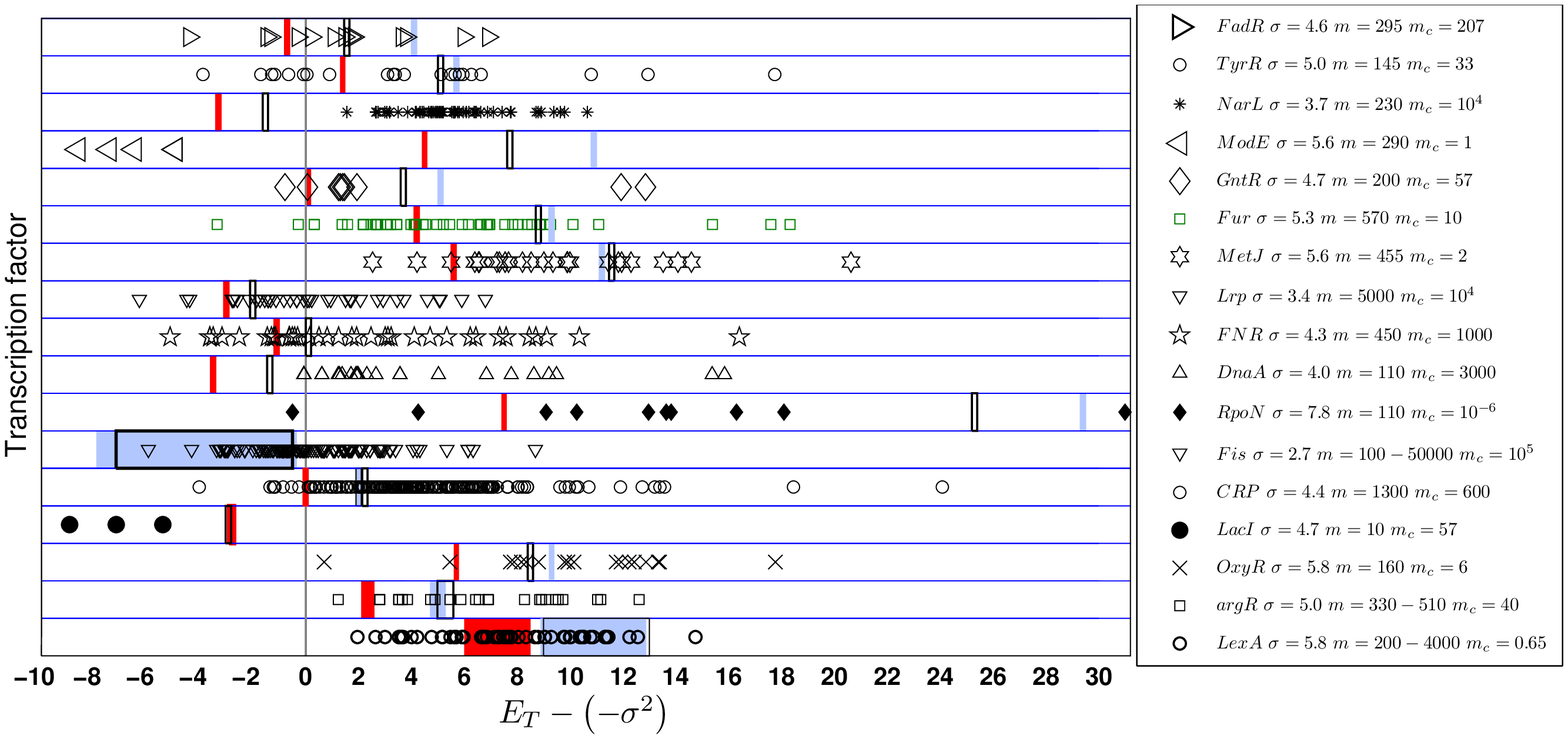}
\caption{A classification of cognate sites of different TFs. For each TF (y-axis) $E_T-\left( -\sigma^2\right) $ is plotted for all its known cognate sites (taken from \cite{Munch2003PRODORIC}). Empty rectangles represent a hypothetical cognate site's energy that is half-occupied if the number of TFs is as specified in the database \cite{AbundanceDatabase} (see legend).
For TFs with $m<m_c$ filled red areas represent the same hypothetical energy, as predicted by Eqs. \eqref{SelfAveSat1}.  When the typical number of the TFs in the cell is larger than $m_c$ the hypothetical cognate sites' energies calculated from the non-steric approximation \eqref{Annealed1} are presented by the red filled area while the results of applying Eq. \eqref{FreezeSat1} are presented by the blue filled areas.
\label{TargetsProteins}}
\end{figure*}

Before presenting a full analysis of the problem it is interesting to consider a case where the different copies of the TFs have steric interactions only on the cognate site. We refer to this as the non-steric approximation. As shown in the SI, here the occupation probability of the cognate site is given by
\begin{equation}
P_{T}^{non-steric}=\frac{1}{1+\frac{N}{m}e^{E_{T}+\frac{\sigma^{2}}{2}}}\;,
\label{Annealed}
\end{equation}
where $E_T$ is the energy of the operator of interest.
Thus, by comparing with Eq. \eqref{Hill} one has
\begin{align}
m_{1/2}&=Ne^{E_T+\frac{\sigma^2}{2}} \;,
\label{Annealed1}\\
n&=1 \;.
\label{Annealed2}
\end{align}
In particular, this implies that without steric interactions there is no enhanced HC and one obtains the naive results.

From the numerical experiment presented before, clearly, the non-steric approximation can fail. In addition to giving a wrong value for $n$ it gives, in some cases, unreasonable values of $m_{1/2}$. Indeed, in Figs. \ref{LexA} and \ref{RpoN} we show numerical results for LexA compared with Eq. \eqref{Annealed}. The values of $m_{1/2}$ differ by more than two orders of magnitude and as stated above the value of $n$ is larger than one. Moreover, the number of LexA TFs in the cell is estimated to be between 200 to 4000, a value much lower than the $m_{1/2}=10^5$ found in the non-steric approximation. A similar disagreement between the numerical results and the non-steric approximation occurs for RpoN  as well as for many other TFs.

To analyze the problem more carefully, taking into account steric interactions, we calculate the occupation probability of the cognate site in two limiting cases (see Methods and SI):

\paragraph{Uncrowded regime:}
In this regime, illustrated in Fig. \ref{Scheme}(a), the number of TFs is smaller than a crossover value defined by the DNA length and the disorder width:
\begin{equation}
m \ll m_c = Ne^{-\frac{\sigma^2}{2}}.
\label{nc1}
\end{equation}
The occupation probability in this regime is given by the non-steric approximation, Eq. \eqref{Annealed}, so that:
\begin{equation}
P_{T}^{uncrowded}=\frac{1}{1+\frac{N}{m}e^{E_{T}+\frac{\sigma^{2}}{2}}}=P_{T}^{non-steric}.
\label{SelfAve}
\end{equation}
We note that the disorder width, $\sigma$, has different values for different TFs with typical values in the range of $2-8$. Thus, the crossover value of the protein's copy number varies from a non-physical $10^{-7}$ (where the non-steric approximation surely fails) to $10^5$ where the non-steric approximation is expected to hold unless  the number of TFs is extremely high.

The above derivation implies that the non-steric approximation (valid in the uncrowded regime) will give a good estimation of the HC and the number of particles at half occupation,
\begin{align}
m_{1/2}&=Ne^{E_T+\frac{\sigma^2}{2}}
\label{SelfAveSat1}\\
n&=1
\label{SelfAveSat2},
\end{align}
only when  $m_{1/2} \ll m_c$. Furthermore, to be in the self averaging regime near  $m \simeq m_{1/2}$  the cognate site of interest, with energy $E_T$,  has to satisfy the condition
\begin{equation}
E_T < -\sigma^2 \;.
\label{StrongTarget}
\end{equation}

\paragraph{Crowded regime:}
In this regime, illustrated in Fig. \ref{Scheme}(b),
\begin{equation}
m \gg m_c = Ne^{-\frac{\sigma^2}{2}}.
\label{nc2}
\end{equation}
and the occupation probability is given by
\begin{equation}
P_{T}^{crowded}=\left\{ 1+\frac{\exp\left[E_{T}+\sigma\sqrt{2\ln\left(\frac{N}{n\sqrt{1+2\sigma^{2}}}\right)}\right]}{\left[\frac{1}{\sigma}\sqrt{2\ln\left(\frac{N}{n\sqrt{1+2\sigma^{2}}}\right)}\right]^{-1}-1}\right\} ^{-1} .
\label{Freeze}
\end{equation}
If close to saturation of the cognate site the system is in the crowded regime, $m_{1/2} \gg m_c$. Then comparing Eqs. \eqref{Freeze} and \eqref{Hill} one can see that 
\begin{align}
m_{1/2}&=\frac{N}{\sqrt{1+2\sigma^2}}e^{ -\frac{W^2\left( e^{-E_T}\sigma^2\right)} { 2\sigma^2} }  
\label{FreezeSat1}\\
n&=\frac{1+W\left( e^{-E_T}\sigma^2\right)}{W^2\left( e^{-E_T}\sigma^2\right)}\sigma^2
\label{FreezeSat2},
\end{align}
where $W$ is the Lambert $W$-function \cite{AS}. The function is well approximated by $W(X) \simeq \ln (X)$ for large $X$.
Note that to be in the crowded regime close to saturation of the cognate site the condition
\begin{equation}
E_T > -\sigma^2
\label{WeakTarget}
\end{equation}
has to be satisfied.

To check our analytical results we plot both the non-steric approximation and the results from the crowded regime in Figs. \ref{LexA} and \ref{RpoN}. The non-steric approximation, \eqref{Annealed}, clearly fails. In contrast Eq. \eqref{Freeze} agrees well with the numerical data (strictly the uncrowded regime result gives a good approximation below $m_c=Ne^{-\frac{\sigma^2}{2}}$). In particular, the HC of the numerical data in the analyzed cases is clearly above one and depends on the disorder width and the cognate site's energy, as demonstrated by Eq. \eqref{Freeze}. 

It is interesting to note that close to saturation the Hill curve (determined by the value of $ m_{1/2} $ and the steepness of $P_T$) is well described by Eqs. \eqref{SelfAveSat1},\eqref{SelfAveSat2},\eqref{FreezeSat1} and \eqref{FreezeSat2}, as presented in Fig. \ref{HalfOccNum}. However, in some cases there is a disagreement between the theory and the numerical results for small values of the protein copy number (see the small $m$ values in Fig. \ref{RpoN}(b) and (c) and the low $ E_T $ values in Fig. \ref{HalfOccNum}(b) and (d)). This disagreement is due to the deviation of the binding energy probability density from a normal distribution at the low energy tail (see Figs. \ref{LexA}(a) and \ref{RpoN}(a)). This effect, which is easy to calculate numerically, increases both the value of $ m_{1/2} $ and $n$ relative to the analytical predictions. 
As shown in Fig. \ref {RpoN}(a) and (b) in some cases this effect may be significant and lead to a very large HC ($n=8.3$ for the argT operon of RpoN). In all the presented cases the non-steric approximation clearly fails by several orders of magnitude.

To examine the validity and relevance of these results to other TFs
we use a database of $17$ known PWMs of TFs of \emph{E. coli}, chosen randomly, and their known binding sites' sequences \cite{Munch2003PRODORIC}. Many of the TFs are present in large copy numbers, larger than their $m_c$ value (see the legend in Fig. \ref{TargetsProteins}). Therefore, the non-steric approximation of the occupation probability, \eqref{Annealed}, does not approximate well the occupation probability of many cognate sequences in the biologically relevant regime. In addition, as one can see in Fig. \ref{TargetsProteins}, the value of $ E_T+\sigma^2 $ for many cognate sites of many proteins is positive. As suggested by Eq. \eqref{WeakTarget}, to describe the occupation probability of such cognate sites Eqs. \eqref{SelfAve}, and \eqref{Freeze} have to be used, while the non-steric approximation, Eq. \eqref{Annealed}, fails. 

It is instructive to calculate the chemical potential or the value of a hypothetical cognate site energy, denoted by $\mu$, which is half-occupied when the number of the TF, $m$, is as it is measured in experiments. The results are shown in Fig. \ref{TargetsProteins}. One can clearly see that the value of $ \mu$ is smaller (larger) than $ -\sigma^2 $ if $m$ is smaller (larger) than $ m_c $, as suggested by conditions \eqref{StrongTarget}, and \eqref{WeakTarget}. Moreover, the non-steric approximation predicts well the value of $ \mu $ for $ m<m_c $ but fails for  $ m>m_c $ predicting much lower values and, therefore predicting very small occupation probability for all the cognate sites with an energy above $\mu$. In contrast, Eqs. \eqref{muSelfAve}, and \eqref{muFreeze} predict well the location of the hypothetical, half-occupied energy for both weak and strong cognate sequences.

\section{Discussion}
The considerations discussed in this paper suggest the existence of a disorder enhanced HC. They provide an estimate of the TF's copy number needed to significantly occupy its cognate sites. This estimate is shown in many cases to be much smaller and more consistent with the existing data than a naive estimate, based on the non-steric approximation.
By analyzing the disordered statistical mechanics problem analytically, we show that two regimes are possible. In the uncrowded regime the number of TFs, $ m $ is much smaller than the crossover value $ m_c $ and steric interactions can be ignored. In contrast, in the crowded regime, $ m \gg m_c $, the steric interactions play an important role and change dramatically the saturation curve, both quantitatively and qualitatively. The crossover value, $ m_c=Ne^{-\sigma^2/2} $, can be much smaller than the DNA length and in many cases is much smaller than the measured number of the TF in the cell.
In addition, for weak cognate sites with a binding energy, $E_T > -\sigma^2$,
the HC is always greater than one, even in the absence of cooperative binding, and in principle unbounded from above.

To understand intuitively this effect we note that in the high disorder regime (where the non-steric approximation fails) the partition function of the system is dominated by a small number of sites with a low energy \cite{Derrida1980}. Therefore, any TF added to the system will immediately bind to one of these low energy sites. Their number can be much smaller than the total number of sites and comparable to the number of TFs. With this in mind it is clear that steric interactions play an important role in this regime and change quantitatively and qualitatively the saturation curve, including a \textit{disorder induced HC} (see Fig. \ref{Scheme} for illustration). 

In the SI we show that this effect persists as long as the proteins spend most of their time in a specific state on the DNA. When this is not the case the value of $m_{1/2}$ grows dramatically making this a rather costly option for many TFs \cite{LBE2009}. In fact, as we show for many of the TFs we consider, data on TF numbers in {\it E. coli} \cite{Dri1994Control} suggests a value of $m_{1/2}$ which does not agree with the partition function which is not dominated by specific conformations of the TF on the DNA. Finally, we show that when naive considerations give a HC of $\overline{n}$ the disorder will lead to a HC which is given by $\overline{n} \cdot n$, with $n$ the {\it disorder enhanced HC} obtained when there is no cooperative binding (see SI).
 
The above results are summarized in Fig. \ref{TargetsProteins} where we plot for several TFs and all their known cognate sites the value of $E_T + \sigma^2$. The regime $E_T+\sigma^2$ smaller/greater than zero corresponds to a without/with disorder enhanced HC. In addition we calculate using data on TF numbers the value of $E_T+\sigma^2$ which would yield $P_T=1/2$ with the typical estimated number of TFs in {\it E. coli}. As can be seen the numbers indicate that a disorder induced HC is likely for a significant fraction of the TFs and their cognate sites. This happens since, as shown in the legend of Fig. \ref{TargetsProteins}, many TFs are present in numbers much larger than $ m_c $ and, therefore, are located in the crowded regime. This study was performed using the measured PWMs of several TFs and can easily be extended to others.
\begin{acknowledgments}
We acknowledge useful comments from A. Horovitz, R. Voituriez, O. Benichou, L. Mirny, E. Braun, A. Sharma and M. Depken. M.S. thanks FOM/NWO for financial support.
\end{acknowledgments}

\section{Methods}
For a cognate site with energy $E_T$ the occupation probability is given by 
\begin{equation}
P_{T}=\frac{1}{1+e^{E_{T}-\mu}}.
\label{Pt}
\end{equation}
The chemical potential, $\mu$, is set by the solution of the equation
\begin{equation}
e^{-F_{ns}+\mu}+\underset{i=1}{\overset{N}{\sum}}\frac{1}{1+e^{E_{i}-\mu}}=m \;.
\label{muEq}
\end{equation}
Here $E_i$ is the binding energy of the TF at location $i=1,2, \ldots N$, with $N$ the number of accessible DNA binding sites, on the DNA. We assume that different copies of the TF exhibit steric interactions, resulting in Fermi-Dirac statistics in Eq. \eqref{muEq}. $F_{ns}$ is the free energy associated with a TF in the solution or in nonspecific conformations on the DNA \cite{BWH81,GMH2002,SM2004,KBBLGBK2004} and, for now, under our assumption (ii) is negligible. In what follows it will be useful to bear in mind that a non-negligible contribution can only reduce the value of $P_T$ and enhance $m_{1/2}$. We discuss its possible contribution in SI. To obtain the occupation probability of the cognate sites of the analyzed TFs we solve numerically Eqs. \eqref{Pt} and \eqref{muEq}.

\section{Supplementary information}

The analysis in the article relied on (i) no inherent cooperativity in the TF, and (ii) a negligible probability to be in non-specific states either on or off the DNA. We now turn to discuss the influence relieving these assumptions.

\subsection {Effects of cooperativity:}
In our study, so far, we ignored the possibility of cooperative effects between distinct molecules of the TF. However, many TFs are active only in their $\bar{n}$-meric form. In this case the HC is usually expected to be close to $\bar{n}$ since the number of active molecules of the TF is proportional to the $\bar{n}$'th power of the concentration of monomers \cite{PKT2009}. One can reformulate the above derivation with $m$ acting as the number of active, $\bar{n}$-meric copies of the TF. In this case the HC according to the arguments given above is $\bar{n} \cdot n$. Thus, a combination of cooperativity and quenched disorder can naturally lead to a high values of the HC.

\subsection {The effect of nonspecific states:}
In our study, so far, we ignored nonspecific states of the protein. These states exist and correspond either to the TF being in the solution or in a nonspecific conformation bound to the DNA \cite{BWH81,GMH2002,SM2004,KBBLGBK2004}. Namely, the nonspecific free energy is given by
\begin{equation}
F_{ns}=\ln\left( e^{-E_{3D}}+Ne^{-E_2}\right),
\end{equation}
where $E_{3D}$ is the free energy of an unbound TF (modeled say by the free-energy of a solution of TFs) and $E_2$ is the energy of the bound protein in the nonspecific conformation which for simplicity we assume to have the same energy for all sites along the DNA (the results are unchanged, being proportional to $N$, in the presence of small disorder in this binding energy with a slight reinterpretation of $E_2$).
Clearly, these nonspecific states can only reduce the occupation probabilities of the cognate sites calculated in Eqs. \eqref{SelfAve} and \eqref{Freeze}. However, as suggested previously the existence of these nonspecific states can be an important component for the dynamics of the search and recognition process carried out by the TF \cite{BWH81,GMH2002,SM2004,Sheinman2011Classes}.

\begin{figure}
\includegraphics[width=\columnwidth]{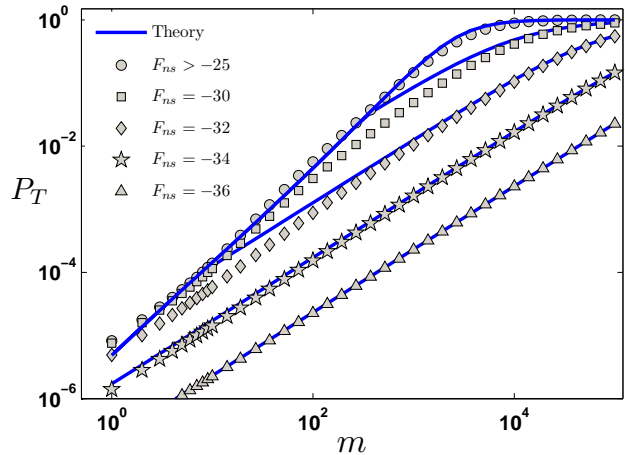}
\caption{The occupation probability of the LexA TF to one of its cognate sequences, the recQ operon, is presented for different values of the nonspecific energy. Different symbols represent the numerical data (see legend) while the solid curves are based on Eq. \eqref{FnsP_T}.
\label{nonspecific}}
\end{figure}

From Eq. \eqref{muEq} one can see that  the nonspecific energy is negligible when
\begin{equation}
F_{ns} \gg \mu+\ln m .
\label{NonSpecificLimit}
\end{equation}
Otherwise a significant fraction of the TFs are in the nonspecific state so that the chemical potential is well approximated by $F_{ns}$. Thus, the occupation probability of the cognate site may be approximated by
\begin{equation}
P_T = \min \left( \frac{1}{e^{E_T-\mu}+1},\frac{1}{\frac{e^{E_T-F_{ns}}}{m}+1} \right) ,
\label{FnsP_T}
\end{equation}
where $ \mu $ is given by Eqs. \eqref{SelfAve}, and \eqref{Freeze}.
In Fig. \ref{nonspecific} we show that indeed this results agree well with the numerical calculations for different values of the nonspecific energy. Thus, condition \eqref{NonSpecificLimit} implies that a half occupation of the cognate site occurs in the freezing regime (so that $n>1$) if
\begin{equation}
F_{ns} \gg E_T+\ln m_{1/2} .
\end{equation}
Since we are not aware of direct measurements of the nonspecific binding it is hard to estimate if this condition holds. We do stress, however, that the experimentally measured numbers of TFs in the cell suggest that it does not play an important role for many TFs.

\begin{figure}
\includegraphics[width=\columnwidth]{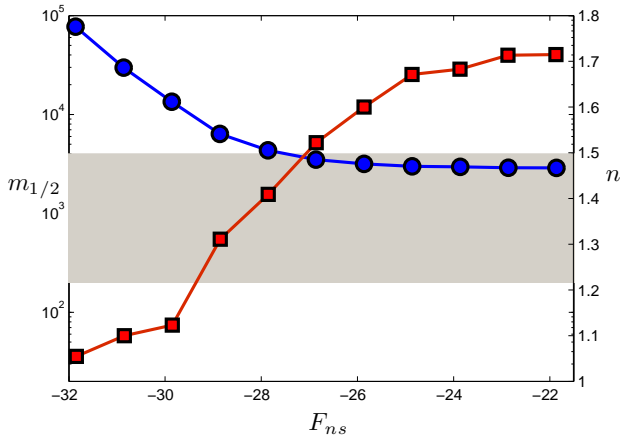}
\caption{The parameters of the fitted Hill curve, Eq. \eqref{Hill}, are presented for the occupation probability of the LexA TF to one of its cognate sequences, the recQ operon as a function of the nonspecific free energy. The circles represent $m_{1/2}$ while the squares represent $n$. The filled gray horizontal area shows the typical range of the LexA copy number in \textit{E.coli} ($200-4000$) \cite{Dri1994Control}.
\label{NonSpecificMeta}}
\end{figure}

In sum, the nonspecific states may be ignored if most of the TFs are mostly bound to the DNA in a specific conformation. Otherwise the chemical potential takes a value which is closer to the nonspecific free energy. This causes the occupation probability to be well approximated by the non-steric approximation which results in a Hill curve with
$
n=1
$
and
$
m_{1/2}=e^{E_T-F_{ns}}.
$
However, in many cases the free energy of the nonspecific states which is needed to bring the system to the self averaging regime is very low. Then the cognate site is significantly occupied only if the number of TFs is much larger than its typical value in the cell. In Fig. \ref{NonSpecificMeta} the parameters of the Hill curve are shown for different values of $ F_{ns} $. One can see that to decrease the HC to one, the value of $ m_{1/2} $ has to be much larger than the typical copy number of the protein. Note, that even in this case the occupation probability might scale as $ m^{n} $ with $ n>1 $ far from the saturation regime, $m_c \ll m \ll \exp\left(F_{ns}-\mu\right) $ (see Fig. \ref{nonspecific}, for an example). In fact the existence of the nonspecific states can totally eliminate the disorder induced HC for all values of $ m $ only if the condition $ F_{ns} \ll \mu+\ln m_c $ is satisfied.

\subsection{Analytical solution}
Assuming a pseudorandom DNA sequence implies that Eq. \eqref{muEq}
is well approximated, for $N \gg 1$, by 
\begin{equation}
\left<\frac{1}{1+e^{E-\mu}}\right>=\int\limits _{-\infty}^{\infty}\frac{\Pr\left(E\right)}{1+e^{E-\mu}}dE=\frac{m}{N} \;,
\label{TheEquation}
\end{equation}
where the binding energy probability density is given by Eq. \eqref{Pr} and the angular brackets are defined by the integral.

First we discuss the non-steric approximation. In this case the Boltzmann statistics takes place of the Fermi-Dirac one everywhere except from the cognate site. Since occupation of the target site is much smaller than $ m $ it can be neglected and the constraint on the chemical potential is given by
\begin{equation}
\left<e^{-E+\mu}\right>=e^{\frac{\sigma^2}{2}+\mu}=\frac{m}{N} \;
\end{equation}
Substituting the solution for $ \mu $ in Eq. \eqref{Pt} one gets the non-steric approximation \eqref{Annealed}.

As is shown in the main text the non-steric approximation clearly fails. We turn now evaluate the integral in Eq.  \eqref{TheEquation} using a saddle point approximation.  The resulting set of equations are
\begin{equation}
\mu=E_*+\ln\left( -\frac{\sigma^2}{E_*}-1\right) 
\label{muSaddlePoint}
\end{equation}
and
\begin{equation}
\frac{e^{-\frac{E_*^2}{2\sigma^2}}}{\sigma^2\sqrt{\frac{1}{\sigma^2}+\frac{1}{E_*+\sigma^2}+\frac{1}{\left( E_*+\sigma^2\right)^2 }}}=\frac{m}{N},
\label{SaddlePoint}
\end{equation}
where $E_*$ is the value of $E$ at the saddle point. Eq. \eqref{SaddlePoint} may be solved self-consistently in two limiting cases:

\paragraph{Uncrowded regime:}
In this regime the saddle point occurs at $E_* \simeq -\sigma^2$, so that Eq. \eqref{SaddlePoint} implies
\begin{equation}
E_*=-\sigma^2\left(1- \frac{m}{N}e^{\sigma^2/2}\right) \;.
\end{equation}
Self-consistency in this regime requires
\begin{equation}
m \ll m_c = Ne^{-\frac{\sigma^2}{2}}.
\label{nc1}
\end{equation}
The chemical potential \eqref{muSaddlePoint} in this limit is given by
\begin{equation}
\mu=-\frac{\sigma^2}{2} -\ln\left(\frac{N}{m} \right) 
\label{muSelfAve}
\end{equation}
and the occupation probability is given by Eq. \eqref{SelfAve}.

\paragraph{Crowded regime:}
In this regime the saddle point satisfies $\left| E_* \right| \ll \sigma^2 $
and Eq. \eqref{SaddlePoint} implies
\begin{equation}
E_*=-\sigma \sqrt{2\ln\left( \frac{N}{m\sqrt{1+2\sigma^2}} \right) }
\end{equation}
Self-consistency requires
\begin{equation}
m \gg m_c = Ne^{-\frac{\sigma^2}{2}}.
\label{nc2}
\end{equation}
The chemical potential in this limit is then given by
\begin{equation}
\mu=-\sigma \sqrt{2\ln\left( \frac{N}{n\sqrt{1+2\sigma^2}} \right) }+\ln\left( \frac{\sigma}{ \sqrt{2\ln\left( \frac{N}{n\sqrt{1+2\sigma^2}} \right) }}-1\right) 
\label{muFreeze}
\end{equation}
and the occupation probability by Eq. \eqref{Freeze}.
\bibliographystyle{apsrev}
\bibliography{BibNoTitle}
\end{document}